\documentclass[dvipdfm,letterpaper,aps,pra,reprint,showpacs,twocolumn,floatfix,amsmath,amssymb]{revtex4-1}
\usepackage{color,graphicx}
\usepackage{dcolumn}
\usepackage{bm}

\newcommand{\Nmax}{{\cal N}_{\text{max}}}
\newcommand{\Tr}{\text{Tr}}
\newcommand{\Gap}{\Delta_{\text{gap}}}
\newcommand{\Nup}{N_{\uparrow}}
\newcommand{\Ndown}{N_{\downarrow}}
\newcommand{\ssec}[1]{\emph{#1}.---}

\begin{document}

\title{Pair condensation in a Finite Trapped Fermi Gas}

\author{C. N. Gilbreth and Y. Alhassid}

\affiliation{Center for Theoretical Physics, Sloane Physics Laboratory, Yale
  University, New Haven, CT 06520, USA}


\begin{abstract}

Superfluidity in the cold atomic two-species Fermi gas system in the unitary limit of infinite scattering length remains incompletely understood. In particular, a pseudogap phase has been proposed  to exist above the superfluid critical temperature. Here we apply the auxiliary-field quantum Monte Carlo method to perform the first \emph{ab initio} calculations of the temperature dependence of three quantities---the energy-staggering pairing gap, the condensate fraction and the heat capacity---in a trapped finite-size cold atom system. As the calculations of the energy-staggering pairing gap require the use of the canonical ensemble, we employ a novel algorithm for the stabilization of particle-number projection that is essential for reaching convergence in the size of the model space. We observe clear signatures of the superfluid phase transition in all three quantities, including a signature of the recently measured lambda peak in the heat capacity, but find no evidence of a pseudogap effect in the energy-staggering pairing gap.

\end{abstract}

\maketitle

Cold atomic Fermi gases have become a central par\-a\-digm in the study of strongly interacting quantum systems and of superfluidity. They provide clean and experimentally tunable interacting systems that can be used as testbeds for the development of many-body methods. The study of cold atomic Fermi gases provides insight into the behavior of such diverse systems as neutron matter, quark matter and high-temperature superconductors~\cite{Chen2005, Gezerlis2008, Baym2008, Bloch2008, Maeda2009}, and they are under wide experimental investigation~\cite{Bloch2008,Giorgini2008}.

Trapped cold atomic Fermi gases can be prepared in a regime in which the scattering length is much larger than any other length scales in the system, allowing the $s$-wave scattering cross section to reach the quantum-mechanical unitary limit. A quantum gas in this unitary limit is strongly interacting, exhibits universal behavior~\cite{Braaten2009,Tan2012}, and lies at a point midway between Bardeen-Cooper-Scrieffer (BCS) and Bose-Einstein condensate (BEC) physics. Understanding the properties of a quantum gas in the unitary limit is thus of broad interest and presents a major theoretical challenge.

Cold atomic Fermi gases exhibit a superfluid phase transition below a critical
temperature $T_c$ because of pairing correlations. A key open question concerns
the nature and extent of pairing above $T_c$ in the unitary limit.  A so-called
``pseudogap phase'' is widely believed to exist above $T_c$ (see \cite{Chen2005}
for an introduction). However, our understanding of this phase remains
incomplete. Measurements of the thermodynamic functions, including the pressure,
compressibility and heat capacity, have shown no signatures of a pseudogap phase
above $T_c$~\cite{Nascimbene2010, Ku2012}, while observations of the spectral
function suggest that pairing correlations do exist at and possibly above
$T_c$~\cite{Stewart2008,Chen2009,Gaebler2010,Perali2011}. Experiments have yet
to measure the pairing gap as a function of temperature across the phase
transition, which could provide a key signature of the phase.

Previous quantum Monte Carlo simulations have varied in their predictions of pseudogap effects in the unitary gas: one group predicted a large pairing gap above $T_c$ in the uniform  gas~\cite{Magierski2009,Magierski2011}, while two other groups did not~\cite{Nascimbene2010,Nascimbene2011,Su2010,Note1}. The former determined the gap $\Delta$ by performing a fit to the maxima of the spectral weight function assuming a BCS-like dispersion relation. We note that most theoretical studies have been carried out for the uniform gas rather than the trapped gas. A local density approximation is used to relate the uniform gas results to the experiments in which the gas is trapped, but it is unclear whether this approximation remains valid in the pseudogap phase.

To date, finite-temperature calculations for cold atomic Fermi gases have
focused on the thermodynamic limit of large numbers of particles. However, the
study of finite-size gases is also of great interest. Examples of strongly-interacting,
finite-size quantum systems with pairing correlations are finite nuclei and  metallic nanoparticles~\cite{Alhassid2013}. Moreover, finite-size gases of strongly interacting cold atoms have recently become experimentally accessible~\cite{Serwane2011}.

Here we apply the auxiliary-field Monte Carlo (AFMC) method to perform \emph{ab initio} calculations of three key quantities across the superfluid phase transition in the trapped cold atom system: (i) the energy-staggering pairing gap $\Gap$, (ii) the condensate fraction and (iii) the heat capacity.  Our calculations are performed for a finite-size ($N=10+10=20$), trapped, two-species (``spin-up'' and ``spin-down'') Fermi gas in the unitary limit.

Our finite-temperature results are the first \emph{ab initio} calculations of quantities (i) and (iii) in any cold atom system, and of quantity (ii) in a finite-size trapped system.  The energy-staggering pairing gap is defined as $\Gap\equiv [ 2 E(\Nup,\Ndown-1) - E(\Nup,\Ndown) - E(\Nup-1,\Ndown-1) ] /2$, where $E(\Nup,\Ndown)$ is the energy of the system with $\Nup$ spin-up and $\Ndown$ spin-down atoms~\cite{Giorgini2008}. It provides a direct and model-independent measure of pairing correlations. However, calculating $\Gap$ requires the finite-temperature Monte Carlo calculations to be performed in the \emph{canonical} ensemble, which for cold atoms has not been achieved to date. Our calculations are made possible by a novel and more efficient algorithm for numerical stabilization in the canonical ensemble, which enables us to achieve convergence in the size of the model space~\cite{Gilbreth2012}.

The study of finite-size systems is qualitatively different from bulk systems, in that finite-size systems do not exhibit phase transitions. However, smooth signatures of phase transitions often still exist. In our calculations, we find clear signatures of the superfluid phase transition in all three quantities studied. In particular, our calculations reveal a signature of the lambda peak in the heat capacity, which was recently observed in experiments~\cite{Ku2012}. However, we find that the energy-staggering pairing gap $\Gap$ does not lead the condensate fraction as the temperature decreases, and thus does not exhibit a signature of a pseudogap phase.

\ssec{Hamiltonian and model space}We consider equal numbers $N_{\uparrow}=N_{\downarrow}=N/2=10$ of two species of fermions interacting at very short range in a spherical harmonic trap with frequency $\omega$. The interaction $V(\bm{r})$ is modeled as a contact interaction $V(\bm{r}) = V_0 \delta (\bm{r})$, which acts only in the $s$-wave channel and therefore allows only particles of different species to interact. We apply the AFMC method in the framework of a configuration-interaction (CI) shell model space spanned by the single-particle eigenstates $|n l m \rangle$ of the harmonic trap ($n$ is the radial quantum number, $l$ is the orbital angular momentum and $m$ is its projection) with energies $\varepsilon_{nl}=(2n+l+3/2)\hbar\omega$. The model space is truncated to $\Nmax$ oscillator shells, i.e., we take only oscillator states with $2n+l \leq \Nmax$. We denote the total number of such states by ${\cal N}_s$. By studying the convergence of observables as a function of $\Nmax$ we determine a cutoff for which the observables are well-converged at all temperatures of interest. The interaction strength $V_0$ for each value of $\Nmax$ is tuned to reproduce the exact ground state energy $E_0$ of the two-particle system ($E_0 = 2 \hbar \omega$ at unitarity).

\ssec{AFMC method}The AFMC method works by applying a Hubbard-Stratonovich (HS) transformation~\cite{Hubbard1959} to the propagator $e^{-\beta \hat{H}}$ (where $\beta$ is the inverse temperature $T$) and evaluating the resulting multidimensional path integral by Monte Carlo methods. We apply the CI shell model formalism commonly used in nuclear physics~\cite{Lang1993,Alhassid1994,Koonin1997,Alhassid2001a} and recently in cold atoms~\cite{Ozen2009}. The propagator is expressed as a path integral
\begin{equation} \label{pathintegral}
  e^{-\beta \hat{H}} \approx \int D[\sigma] G_\sigma \hat U_\sigma \;,
\end{equation}
where the integration is over auxiliary fields $\sigma(\tau)$ that depend on the imaginary time $\tau$ ($0\leq \tau \leq \beta$), $\hat U_\sigma$ is the many-particle propagator for a system of noninteracting fermions moving in external fields $\sigma(\tau)$, and $G_\sigma$ is a Gaussian weight. To facilitate numerical evaluation the imaginary time is discretized into a finite number $N_\tau$ of points $\tau_n = n \Delta \beta$, where $\Delta \beta = \beta/N_\tau$. The calculations become exact in the limit $\Delta \beta \rightarrow 0$. We find that observables scale linearly versus $\Delta\beta$ for $\Delta \beta \leq 1/32$ and take the limit $\Delta \beta \rightarrow 0$ by a linear extrapolation.

\ssec{Observables}In AFMC, a set of configurations $\sigma_k$ are sampled according to the distribution $W_\sigma=G_\sigma |\Tr \hat U_\sigma|$ using the Metropolis algorithm~\cite{Koonin1997}. Thermal expectation values of observables are then calculated from \begin{equation}
  \frac{\Tr(\mathcal{\hat O} e^{-\beta \hat H })} {\Tr(e^{-\beta \hat H })}
  \approx \frac{\sum_k \langle \mathcal{\hat O} \rangle_{\sigma_k}
    \Phi_{\sigma_k}}{\sum_k \Phi_{\sigma_k}} \;,
  \label{expectation}
\end{equation}
where $\langle \mathcal{\hat O} \rangle_\sigma = \Tr(\mathcal{\hat O} \hat U_\sigma)/\Tr \hat U_\sigma$ is the expectation of an operator $\mathcal{\hat{O}}$ at a given configuration $\sigma_k$ of the auxiliary fields and $\Phi_\sigma \equiv \Tr(\hat U_\sigma) / |\Tr \hat U_\sigma|$ is the Monte Carlo sign.

Unlike most other AFMC calculations, we calculate the traces in (\ref{expectation}) and in the weight function $W_\sigma$ at fixed particle numbers $N_{\uparrow}, N_{\downarrow}$, i.e., in the canonical ensemble. This is accomplished by using an exact particle-number projection via a discrete Fourier transform~\cite{Ormand1994}.

\ssec{Monte Carlo sign}In general the sign $\Phi_\sigma$ is a sample-dependent complex phase. If the fluctuations in $\Phi_\sigma$ become comparable to its average value, the statistical error in (\ref{expectation}) becomes excessively large, giving rise to the so-called Monte Carlo sign problem. However, attractive contact interactions are known to have good sign ($\Phi_\sigma = 1$ for all $\sigma$) in the grand-canonical ensemble~\cite{Bulgac2006}, thereby avoiding the problem. This is also the case in our canonical ensemble calculations, and can be seen as follows. The sign is determined by particle-projected two-species trace \begin{equation}\label{trace}
  \Tr_{N_{\uparrow},N_{\downarrow}} \hat U_\sigma = \left(\Tr_{N_{\uparrow}}\hat
  U^\uparrow_\sigma\right) \left(\Tr_{N_{\downarrow}}\hat
  U^\downarrow_\sigma\right)\;,
\end{equation}
where $\hat U^\uparrow_\sigma$ ($\hat U^\downarrow_\sigma$) are the auxiliary-field propagators for spin-up (spin-down) particles.  For an attractive contact interaction, $\hat U^\uparrow_\sigma$ and $\hat U^\downarrow_\sigma$ are time-reversal invariant, and both particle-projected traces on the right-hand side of (\ref{trace}) are real. For the spin-balanced system $N_{\uparrow}=N_{\downarrow}$, we have $\Tr_{N_{\uparrow},N_{\downarrow}} \hat U_\sigma = (\Tr_{N_{\uparrow}}\hat U^\uparrow_\sigma)^2>0$ so the canonical sign is also 1.

\ssec{Accuracy and convergence}The accuracy of our calculations is determined by two factors. First, the contact interaction is not an exact representation of the usual regularized delta function $V(r) = \delta({\bm r}) \partial / \partial r$ used to model cold atoms~\cite{Esry1999}. We checked its accuracy against the known analytic solution of the three-particle system and found the energy to be accurate within $1\%$.

\begin{figure}
  \includegraphics[width=0.42\textwidth]{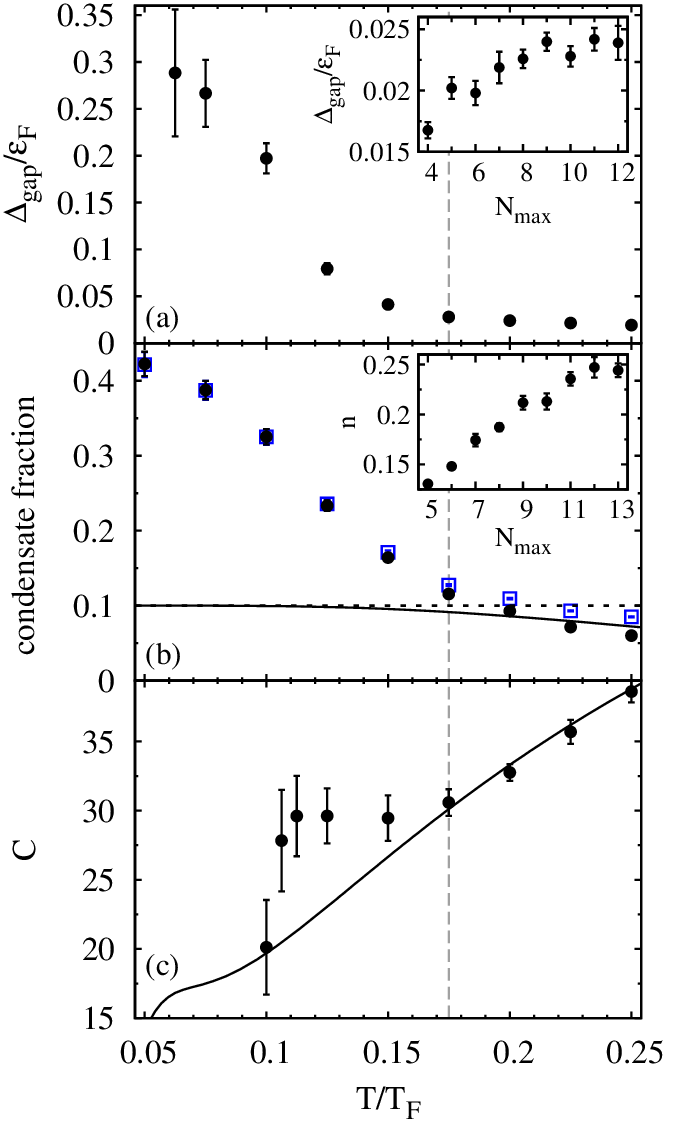}
  \caption{Signatures of the superfluid phase transition in the trapped spin-balanced $N=20$ atom system ($\Nup=\Ndown=10$). (a) AFMC results for the pairing gap $\Gap$ vs. temperature.  The inset shows the convergence of the gap as a function of the number of oscillator shells $\Nmax$ at $T/T_F = 0.2$. (b) Condensate fraction vs. temperature. Solid circles: the scaled occupation $n_0$ of the $T=0$ pair wavefunction. Open squares: the largest scaled eigenvalue $n$ of the $L=0$ pair correlation matrix (\ref{pair-correlation}). Solid line: noninteracting result. The dashed
    horizontal line indicates the noninteracting upper bound of $1/(N/2)$. The inset shows the convergence of $n$ as a function of $\Nmax$ at $T/T_F = 0.125$. (c) Heat capacity vs. temperature. The AFMC results (solid circles) are compared with the heat capacity of noninteracting fermions in the trap (solid line). The vertical dashed line in the three panels corresponds to a temperature of $T/T_F=0.175$ (see text). \label{allvsT}}
\end{figure}

Second, for the calculations to be accurate, the model space must be large enough to account for both interaction effects and thermal excitations.  We have accounted for this by using model spaces that are sufficiently large for the observables of interest to be well-converged for the relevant particle number and temperatures. We demonstrate the convergence in the insets to Figs.~\ref{allvsT}(a) and \ref{allvsT}(b); the pairing gap $\Delta_{\rm gap}$ is well-converged by $\Nmax=9$, and the condensate fraction $n$ by $\Nmax=11$. We used these values of $\Nmax$ in our calculations, along with $\Nmax=11$ for the heat capacity (which is also well-converged).

\ssec{Algorithmic improvements}
In AFMC the thermal propagator $\hat{U}_{\sigma}$ for a given set of auxiliary fields $\sigma$ is represented as a chain of matrix products,
\begin{equation}\label{chain}
U_{\sigma} = U_{\sigma}^{(N_t)} \cdots U_{\sigma}^{(1)}\,,
\end{equation}
where $N_t = \beta/{\Delta}\beta$ is the number of time slices and $U_{\sigma}$ is the matrix representation of $\hat{U}_{\sigma}$ in the basis of single-particle states. As the
temperature decreases, $N_t$ becomes large and the product~\eqref{chain} becomes
ill-conditioned. A well-known method to  stabilize such products is to apply a
matrix decomposition~\cite{LohJr1992},
\begin{equation} \label{decomp}
U_{\sigma} = A D B\,,
\end{equation}
where the matrices $A$ and $B$ are well-conditioned and $D$ is diagonal with
positive entries (for instance, $ADB$ could be a singular value
decomposition). One initially applies the decomposition~\eqref{decomp} to the
first time slice $U_{\sigma}^{(1)}$, then carefully updates it as the product $U_{\sigma}^{(N_t)} \cdots
U_{\sigma}^{(1)}$ is constructed.

To calculate the partition function in the canonical ensemble, a Fourier
transform method is most convenient~\cite{Ormand1994}. This requires calculating ${\cal N}_s$
grand-canonical partition functions, where ${\cal N}_s$ is the number of
single-particle states:
\begin{equation}\label{FT}
\Tr_{p}\,\hat{U}_{\sigma}\, = \frac{1}{{\cal N}_s} \sum_{m=1}^{{\cal N}_s} e^{i \varphi_m p} \det (1+U_{\sigma} e^{-i \varphi_m})\,.
\end{equation}
In Eq.~\eqref{FT}, $\Tr_{p}\,\hat{U}_{\sigma}$ is the canonical partition function
for $p=N_{\uparrow}$ or $N_{\downarrow}$ particles of a single species.  A
straightforward way to calculate~\eqref{FT} stably from the decomposition
$U=ADB$ is to apply the usual method for the grand-canonical
ensemble~\cite{LohJr1992} to each term in the Fourier
sum~\cite{Alhassid2008}. However, this method involves a matrix decomposition
for each term and consequently scales as ${\cal O}({\cal N}_s^4)$. Calculations with
this method are too slow to reach sufficiently large model spaces for
convergence in the system we study here.

Our new approach is to diagonalize the cyclic permutation $D B A$ of the
decomposition~\eqref{decomp} before computing the Fourier
transform~\eqref{FT}. The matrix $U = A D B$ cannot simply be multiplied out and
diagonalized, as this would involve a severe loss of information about the
intermediate numerical scales contained in $U$. However, the permutation $D B A$
can be constructed stably, and we find that, even though it is highly
ill-conditioned, it can also be diagonalized stably (in ${\cal O}({\cal N}_s^3)$
time). The eigenvectors and eigenvalues of $U$ can be easily recovered from
those of $D B A$, permitting the Fourier transform to be efficiently
computed. Overall, this method scales as ${\cal O}({\cal N}_s^3)$ and permits
our canonical-ensemble calculations to reach model spaces much larger than what
would otherwise be possible. In particular, $\Nmax=11$ contains ${\cal N}_s=364$
single-particle states and to our knowledge is the largest model space used to
date for canonical-ensemble AFMC calculations.

\ssec{Pairing gap}We calculated the pairing gap $\Gap$ by particle-number reprojection~\cite{Alhassid1999}, in which only one Metropolis walk is needed to calculate all three energies $E(N_{\uparrow}-1,N_{\downarrow}-1)$, $E(N_{\uparrow},N_{\downarrow}-1)$ and $E(N_{\uparrow},N_{\downarrow})$ contributing to $\Gap$. We sampled the fields according to the distribution $G_\sigma |\Tr_{N_{\uparrow},N_{\downarrow}-1} \hat U_\sigma|$.

Our result for the pairing gap is shown in Fig.~\ref{allvsT}(a) as a function of temperature. Here $\Gap$ is plotted in units of $\varepsilon_F \equiv 4.0\,\hbar\omega$~\cite{Note2} and the temperature is given in units of $T_F\equiv \varepsilon_F/k_B$, where $k_B$ is the Boltzmann constant.  At high temperature, the finite size of the system causes $\Gap$ to remain slightly above zero. As the temperature decreases, the gap begins to depart from its high-temperature behavior at around $T/T_F \approx 0.175$; it then rises rapidly before saturating at approximately $T/T_F=0.07$, providing a clear signature of a superfluid phase transition. We estimate the zero-temperature pairing gap by averaging the values at the lowest two temperatures, obtaining $\Gap = 0.271(32) \varepsilon_F$. This result is consistent with fixed-node diffusion Monte Carlo ~\cite{Blume2007} and density functional theory~\cite{Bulgac2007a} calculations.

It has been predicted that at zero temperature the energy-staggering pairing gap in a trap will be suppressed compared to its value in the homogeneous system, since in a trap unpaired particles can locate themselves in the edge of the cloud, where the energy required to add an extra particle to the system is smaller~\cite{Son2007}. More precisely, $\Gap$ was predicted to scale at zero temperature as $\Gap \propto (N+1)^{1/9}\hbar\omega$. Using $\varepsilon_F = (3N)^{1/3} \hbar\omega$~\cite{Giorgini2008}, one therefore expects $\Gap / \varepsilon_F \propto N^{-2/9}$ for large $N$. We compare this scaling with our results by computing the pairing gap at low temperature for $N = 10$ ($N_{\uparrow}=N_{\downarrow}=5$). We find $\Gap(5,5)=0.294(21)$, so that $\Gap(10,10)/\Gap(5,5) = 0.92(13)$, surprisingly close (for such small $N$) to the expected value of $0.866$, although results at larger values of $N$ would be required for a conclusive comparison.

\ssec{Condensate fraction}Although condensation is inherently a macroscopic
phenomenon, a reasonable definition of a condensate fraction can be made for
finite-size systems.
For uniform systems, there is an equivalence between off-diagonal long-range
order (ODLRO) and the existence of a large eigenvalue in the two-body density
matrix~\cite{Yang1962}. Thus, the existence of a large eigenvalue in this matrix
is connected with the presence of superfluidity. In the trapped system, the
orbital angular momentum $L$ is conserved, and we calculate the pair correlation
matrix~\cite{Alhassid1996} \begin{equation}\label{pair-correlation} C_{L}(a b,c
  d) \equiv \langle A^{\dagger}_{L M \uparrow \downarrow}(a b) A_{L M \uparrow
    \downarrow}(c d) \rangle \;,
\end{equation}
where $\langle \cdot \rangle$ denotes a thermal expectation value as in Eq.~(\ref{expectation}). Here $A^\dagger_{LM \uparrow \downarrow}(ab)$ is a pair creation operator for two particles in orbitals $a = (n_a,l_a)$ and $b=(n_b,l_b)$ coupled to total angular momentum $L$ with magnetic quantum number $M$, with the first particle being of the $\uparrow$ species and the second the $\downarrow$ species.  The expectation values on the right-hand side of (\ref{pair-correlation}) are independent of $M$ because of rotational invariance. Up to a $(2L+1)$-fold degeneracy, the eigenvalues of $C_{L}(a b,c d)$ are exactly those of the matrix $\langle a^\dagger_{a m_a \uparrow} a^\dagger_{b m_b \downarrow} a_{d m_d \downarrow} a_{c m_c \uparrow} \rangle$, which is the the orbital-space form of the density matrix studied in the uniform gas (via ODLRO)~\cite{Bulgac2006,Astrakharchik2005,Salasnich2005,Oritz2005}. We find that the largest eigenvalue always occurs for $L=0$.

As the temperature decreases, the maximum eigenvalue $\lambda_{\rm max}$ becomes much larger than all others and satisfies constraints which allow for two natural definitions of a condensate fraction. To make these definitions, we note that the corresponding eigenvector $\varphi(ab)$ defines a pair creation operator $B^{\dagger} = \sum_{a,b} \varphi^\ast(ab) A^\dagger_{0 0 \uparrow \downarrow}(a b)$ which satisfies $\langle B^\dagger B \rangle = \lambda_{\rm max}$, indicating that $\lambda_{\rm  max}$ is the occupation of a two-body wavefunction. For free fermions, $0 \leq \lambda_{\rm max} \leq 1$, so no more than two fermions can occupy a paired state. However, in the presence of interactions $\lambda_{\rm max}$ can exceed $1$, effectively allowing fermion pairs to ``condense''. In this case one can show that
 $\lambda_{\rm max}$
satisfies $0 \leq \lambda_{\rm max} \leq B({\cal N}_s,N)$, where
\begin{equation}
  B({\cal N}_s,N) = N ({\cal N}_s - N/2 + 1) / 2 {\cal N}_s \leq N/2\,,
\end{equation}
and ${\cal N}_s$ is the number of single-particle orbitals for a single species~\cite{Note3}.  In the limit of large ${\cal N}_s$, $B({\cal N}_s,N)$ approaches $N/2$. For $\Nmax=11$ oscillator shells ${\cal N}_s=364$, and $B({\cal N}_s,N)=9.75$ is quite close to $N/2=10$. Moreover, $\lambda_{\rm max}=B({\cal N}_s,N)$ can be achieved for a particular many-body wavefunction (which is not necessarily an eigenstate of the system). We may therefore define a condensate fraction by 
\begin{equation}
n \equiv \lambda_{\text{max}}/(N/2)\,,
\end{equation}
where $0 \leq n \leq 1$.

We can also define a condensate fraction in terms of the scaled occupation of the zero-temperature pair $B_0^\dagger \equiv B^\dagger(T=0)$, i.e., 
\begin{equation}
n_0 \equiv \langle B_0^\dagger B_0 \rangle / (N/2)\,.
\end{equation}
In contrast to $n$, $n_0$ measures the fraction of fully condensed pairs at finite temperature, ignoring contributions from paired particles not occupying the $T=0$ pair wavefunction. In general $n_0(T) \leq n(T)$ and equality holds in the limit $T=0$.

Fig.~\ref{allvsT}(b) shows our result for $n_0$ and $n$ as a function of temperature. The two condensate fractions are nearly identical for temperatures below $T/T_F=0.175$, while $n_0 < n$ at higher temperatures. For comparison, we plot $n$ for a noninteracting gas (solid line) and its largest possible value of $2/N$ corresponding to $\langle B^\dagger B \rangle = 1$ (dashed line). In both cases we see a rapid increase of the condensate fraction below $T/T_F \approx 0.175$, signifying the onset of a superfluid state.

The condition $\langle B_0^\dagger B_0 \rangle > 1$ provides a definite upper bound $T_{\rm ub}$ in temperature for condensation to be present, since condensation requires the pair wavefunction to be occupied by more than one Fermion pair. In our system, this yields $T_{\rm ub}/T_F \approx 0.175$, a temperature similar to the transition temperature scales for the pairing gap and the heat capacity (see below for the latter).

\ssec{Pseudogap effects}An effective way to assess pseudogap effects in this system is to compare the temperature dependence of the condensate fraction with the temperature dependence of the pairing gap. In the pseudogap phase it is expected that preformed pairs of attractive interacting fermions would cause the pairing gap to lead the condensate as the temperature decreases towards $T_c$.  Comparing Figs.~1(a) and 1(b) indicates that this does not occur for $\Gap$. First, $\Gap(T_{\rm ub}) = 0.028(2)$, an insignificant fraction of its $T=0$ value. Second, below $T/T_F\approx0.175$, the condensate fractions $n$ and $n_0$ grow simultaneously with $\Gap$. Thus, we conclude that in the $N=20$ finite-size system, $\Gap$ does not display any pseudogap effects.

Similar to our results, the $T$-matrix calculations of Ref.~\cite{Tsuchiya2011} for a trapped gas found a pseudogap temperature only slightly higher than the critical temperature for the unitary trapped gas. In contrast, Monte Carlo calculations for a \emph{uniform} gas found a nonzero gap $\Delta$ at temperatures significantly above $T_c$~\cite{Magierski2009,Magierski2011}. In those calculations $\Delta$ was extracted from the spectral weight function (by fitting a BCS-like dispersion to its maxima) rather than from energy staggering.

It would be interesting to use our method to study the finite-temperature behavior of the pairing gap and condensate fraction as a function of $N$. At zero temperature, fixed-node quantum Monte Carlo and density-functional calculations predict that finite-size effects become negligible for more than 50 particles~\cite{Forbes2011}, and lattice Monte Carlo computations have found that shell effects persist at the 2\% level above 40 particles~\cite{Endres2011}.

\ssec{Heat capacity}Fig.~\ref{allvsT}(c) shows the heat capacity $C = dE/dT$ as a function of temperature. We calculated $C$ by numerically differentiating the energy inside the path integral; this method takes into account correlated errors and greatly reduces the statistical error compared to differentiation after calculating $E(T)$~\cite{Alhassid2001}. At high temperatures, the heat capacity agrees with that of a noninteracting trapped Fermi gas (solid line). As the temperature decreases below $T/T_F=0.175$, it begins to deviate from its non-interacting limit, remaining elevated until dropping rapidly at $T/T_F=0.11$---a smoothed but obvious signature of the lambda peak observed recently in the experiments~\cite{Ku2012}. This smoothed behavior is expected for finite-size systems, where sharp phase
transitions do not occur, but strong signatures may still exist~\cite{Alhassid2013}. The temperature at which this structure emerges, $T/T_F \approx 0.175$, is commensurate with the rise of $\Gap$ and the temperature at which the condensate fraction $n$ exceeds its noninteracting limit.

In conclusion,  using a quantum Monte Carlo method we have completed the first \emph{ab initio} calculations of the energy-staggering pairing gap $\Gap$, condensate fraction, and heat capacity as a function of temperature for a finite-size, unpolarized system of trapped atoms at unitarity. The calculation of $\Gap$ requires the use of the canonical ensemble in which the number of atoms is fixed, and was made possible by a novel algorithm for stabilization of particle-number projection.  We have identified clear signatures of the superfluid phase transition in each of the three studied quantities. However, in this finite-size trapped system, $\Gap$ does not lead the condensate fraction as temperature decreases and thus does not exhibit any signature of a pseudogap phase.

The condensate fraction has been probed in experiments of larger systems by rapidly ramping the gas from the unitary regime to the BEC regime to transform fermion pairs into tightly bound molecules (see, e.g., Refs.~\cite{Regal2004} and [9]). It might be possible to apply a similar method in experiments on finite-size systems such as those of Ref.~\cite{Serwane2011}, where finer control over particle number may also allow measurement of the energy-staggering gap.

The same AFMC method used here can also be applied on either side of the BEC-BCS crossover, and could potentially be used to study finite-size scaling.

We thank A.~Mukherjee for referring us to Ref.~\cite{Yang1962}, and M.~M.~Forbes and K.~R.~A.~Hazzard for useful discussions. This work was supported in part by the Department of Energy grant DE-FG-0291-ER-40608. Computational cycles were provided by the facilities of the Yale University Faculty of Arts and Sciences High Performance Computing Center and by the NERSC high performance computing facility at LBL.


\end{document}